\documentclass[doc,12pt]{apa}

\usepackage[T1]{fontenc}
\usepackage[utf8]{inputenc}%
\usepackage{apacite}
\usepackage{graphicx}
\usepackage{booktabs} 
\usepackage{amssymb}

\title{\Huge\bf Spatial frequency processing in the central and peripheral visual field during scene viewing} 
\author{Anke Cajar, Ralf Engbert, and Jochen Laubrock}
\affiliation{Department of Psychology, University of Potsdam, Germany}

\rightheader{Gaze-contingent spatial frequency filtering}

\begin{document}
\maketitle

\vspace{2cm}
\vspace{\fill}

\ \\ Adress correspondence to:

\ \\ Anke Cajar\\
Department Psychologie\\ 
Universität Potsdam\\
Karl-Liebknecht-Str. 24-25\\ 
14476 Potsdam, Germany\\ \\
Email: cajar@uni-potsdam.de\\
Phone: +493319772127\\ 
Fax: +493319772793\\

\newpage
\section{Abstract}
Visuospatial attention and gaze control depend on the interaction of foveal and peripheral processing. The foveal and peripheral regions of the visual field are differentially sensitive to parts of the spatial-frequency spectrum. In two experiments, we investigated how the selective attenuation of spatial frequencies in the central or the peripheral visual field affects eye-movement behavior during real-world scene viewing. Gaze-contingent low-pass or high-pass filters with varying filter levels (i.e., cutoff frequencies; Experiment 1) or filter sizes (Experiment 2) were applied. Compared to unfiltered control conditions, mean fixation durations increased most with central high-pass and peripheral low-pass filtering. Increasing filter size prolonged fixation durations with peripheral filtering, but not with central filtering. Increasing filter level prolonged fixation durations with low-pass filtering, but not with high-pass filtering. These effects indicate that fixation durations are not always longer under conditions of increased processing difficulty. Saccade amplitudes largely adapted to processing difficulty: amplitudes increased with central filtering and decreased with peripheral filtering; the effects strengthened with increasing filter size and filter level. In addition, we observed a trade-off between saccade timing and saccadic selection, since saccade amplitudes were modulated when fixation durations were unaffected by the experimental manipulations. We conclude that interactions of perception and gaze control are highly sensitive to experimental manipulations of input images as long as the residual information can still be accessed for gaze control.

\newpage

\section{Introduction}

Why do we move our eyes? Due to sensory and cognitive limitations, high-acuity vision is restricted to the central 2$^\circ$ of the visual field, the fovea, whereas the visual periphery is rather blurry \cite{Strasburger.JVis.2011,Wertheim.ZPsycholPhysiolSinnesorgane.1894}. As a consequence, high-velocity saccades shift the gaze about three times each second to bring regions of interest from the low-resolution periphery into the fovea for closer inspection. Two tasks are accomplished during the following fixation: fine-grained foveal information is analyzed to identify objects and details, and coarse-grained peripheral information is analyzed to select the next saccade target among competing regions of interest. Thus, visual information in the central and the peripheral visual field serve different tasks \cite{Gilchrist.INBOOK.2011}.

The present study investigates how the two tasks of foveal analysis and peripheral selection are accomplished during real-world scene viewing when fine-grained or coarse-grained information is selectively attenuated in the central or the peripheral visual field. Inherently this also sheds light on the question to what degree central and peripheral vision contribute to spatial and temporal aspects of eye-movement behavior. The issue can be tackled by attenuating high or low spatial frequencies in the central or the peripheral visual field. High spatial frequencies provide the fine-grained information of an image and low spatial frequencies provide the coarse-grained information of an image. High-pass filters preserve high spatial frequencies and attenuate low spatial frequencies; with low-pass filters, it is vice versa. Information can be selectively altered in either the central or the peripheral part of the visual field by applying a gaze-contingent window of arbitrary size that moves with the current gaze position of the viewer in real-time during scene inspection \cite{McConkie.PerceptPsychophys.1975,Rayner.Science.1979}. Spatial frequencies are filtered either inside or outside the gaze-contingent window with central or peripheral filtering respectively, while the other region of the scene remains unchanged.

Previous research on this topic is rather scant and has mostly been focused on the effects of peripheral low-pass filtering. Corresponding studies indicate that spatial-frequency filtering impairs scene processing, as viewers' performances in several tasks decrease with filtering. For example, when searching for objects in scenes, search accuracy decreases and search times increase with peripheral as well as with central low-pass filtering; these effects get stronger as filter level and filter size increase \cite{Loschky.JExpPsycholAppl.2002,Nuthmann.JExpPsycholHuman.2014}. Furthermore, the probability to detect target stimuli in low-pass or high-pass filtered scene regions decreases and response times for detected targets increase \cite{Cajar.JVis.2016}. Central low-pass filtering has also been shown to decrease response accuracy to memory questions about scenes \cite{Cajar.JVis.2016}. These findings suggest that the processing difficulty of a scene increases with spatial-frequency filtering, and increases more the larger or stronger the filter gets.

In agreement with the decrease in task performance, eye-movement behavior has been reported to deviate progressively from viewing behavior in unfiltered scenes as spatial-frequency filtering increases processing difficulty. Studies consistently show that viewers prefer unfiltered scene regions as saccade targets. Peripheral filtering shortens mean saccade amplitudes, since viewers tend to keep their gaze inside the unfiltered central region and avoid longer saccades to the filtered periphery \cite{Cajar.JVis.2016,Foulsham.AttenPerceptPsychophys.2011,Laubrock.JVis.2013,Loschky.JExpPsycholAppl.2002,Loschky.VisCogn.2005,Nuthmann.VisCogn.2013,Nuthmann.JExpPsycholHuman.2014,Shioiri.Perception.1989}. Central filtering, on the other hand, lengthens mean saccade amplitudes, since viewers tend to place fewer saccades inside the filtered center and make more long saccades to the periphery \cite{Cajar.JVis.2016,Laubrock.JVis.2013,Nuthmann.JExpPsycholHuman.2014}. With both central and peripheral low-pass filtering, the effects get larger with increasing filter size \cite{Loschky.JExpPsycholAppl.2002,Nuthmann.VisCogn.2013,Nuthmann.JExpPsycholHuman.2014} and filter level \cite{Loschky.JExpPsycholAppl.2002}. Thus, saccadic selection is modulated more and more as processing difficulty increases. It has been shown recently that these changes in saccade amplitudes go along with corresponding changes in visuospatial attention \cite{Cajar.JVis.2016}.

Fixation duration also varies with visual-cognitive processing and usually increases as the acquisition of information from the scene becomes more difficult \cite{Henderson.TrendsCognSci.2003,Nuthmann.PsycholRev.2010}. Thus, studies show that fixation durations increase with spatial-frequency filtering of the entire scene \cite{Glaholt.AttenPerceptPsychophys.2013,Henderson.VisCogn.2014,Mannan.SpatVis.1995} as well as with central low-pass filtering \cite{Cajar.JVis.2016,Nuthmann.JExpPsycholHuman.2014} and with peripheral low-pass filtering \cite{Cajar.JVis.2016,Laubrock.JVis.2013,Loschky.JExpPsycholAppl.2002,Loschky.VisCogn.2005,Nuthmann.VisCogn.2013,Nuthmann.JExpPsycholHuman.2014,Parkhurst.INBOOK.2000,vanDiepen.Perception.1998}. Fixations also increasingly prolong with increasing low-pass filter size \cite{Nuthmann.VisCogn.2013,Nuthmann.JExpPsycholHuman.2014,Parkhurst.INBOOK.2000}. However, Loschky and colleagues found that increasing filter level with detectable peripheral low-pass filtering hardly affected fixation durations \cite{Loschky.JExpPsycholAppl.2002,Loschky.VisCogn.2005}. In summary, previous research suggests that eye-movement behavior is increasingly modulated as visual-cognitive processing difficulty increases due to spatial-frequency filtering. 

In contrast, we recently found evidence in two studies \cite{Cajar.JVis.2016,Laubrock.JVis.2013} that fixation durations are not always longer under conditions of increased processing difficulty. In both studies, high-pass filters or low-pass filters were applied to either the central or the peripheral part of the visual field during the viewing of color \cite{Laubrock.JVis.2013} or grayscale \cite{Cajar.JVis.2016} real-world scenes. We assumed that scene processing would be most difficult with central low-pass and peripheral high-pass filtering, as these conditions strongly attenuate the critical spatial frequencies for foveal analysis (high spatial frequencies) and peripheral target selection (low spatial frequencies) respectively. In both studies, however, mean fixation durations increased most with central high-pass and peripheral low-pass filtering, which were expected to be less disruptive for processing. Central low-pass and peripheral high-pass filtering involved shorter mean fixation durations, often similar to the mean fixation duration in the unfiltered control condition. The results suggest that viewers invested more processing time when the information left after filtering was useful enough to accomplish the task at hand (foveal analysis, peripheral selection) in a reasonable amount of time; when visual-cognitive processing became too difficult to make an investment of more processing time worthwhile default timing, that is, stimulus-independent random timing of saccades was adapted. To account for these effects, we developed a computational model in which fixation durations are controlled by the dynamical interaction of foveal and peripheral processing \cite{Laubrock.JVis.2013}. The model assumes that foveal and peripheral information processing evolve in parallel and independently from one another during fixation, a notion that was recently corroborated by an experimental study \cite{Ludwig.PNAS.2014}.

\subsection{The present study}
Most studies on the effects of gaze-contingent spatial-frequency filtering on eye movements during scene viewing applied peripheral low-pass filters. There is only little research on the effects of central filtering \cite{Nuthmann.JExpPsycholHuman.2014} and high-pass filtering \cite{vanDiepen.Perception.1998} on eye-movement behavior. Our own studies \cite{Cajar.JVis.2016,Laubrock.JVis.2013} were the first to investigate the effects of central and peripheral high-pass and low-pass filtering within the same experiment. Moreover, to our knowledge, there is no study on the effects of varying filter level and filter size with high-pass filtering. The investigation of the latter is interesting on its own. In addition, the aforementioned effects on fixation durations in our previous studies with varying filter type (low-pass/high-pass) in the central or peripheral visual field are striking---they raise the question how fixation durations adapt to processing difficulty due to spatial frequency filters of varying filter level and filter size. The present study investigated this question in two experiments.

In both experiments, participants inspected real-world scenes in preparation for a memory task while high or low spatial frequencies were filtered either in the central or the peripheral visual field. Gaze-contingent filtering was compared with control conditions that presented scenes either unfiltered or entirely low-pass or high-pass filtered. In Experiment 1, the level of filtering (i.e., the cutoff frequency) varied between trials using weak, moderate, or strong high-pass or low-pass filters. Processing difficulty was assumed to increase from weak to strong filters. In Experiment 2, filter level was constant, but the size of the filter (i.e., the size of the gaze-contingent window) varied---the filter either subtended a small, medium, or large region of the central or the peripheral visual field. Processing difficulty was assumed to increase from small to large filters. The experiments tested for the effects of filtering on task performance, fixation durations, and saccade amplitudes. 

For both experiments, we expected saccade amplitudes to increasingly deviate from normal viewing behavior with increasing processing difficulty. Compared with unfiltered scene viewing, amplitudes were expected to increase with central filtering and decrease with peripheral filtering, particularly when critical spatial frequencies were attenuated (i.e, with central low-pass filtering and peripheral high-pass filtering). These effects were expected to grow with increasing filter level (Experiment 1) or filter size (Experiment 2). For fixation durations, we expected an increase as long as potentially more useful spatial frequencies were preserved and information uptake was relatively easier, that is, with central high-pass and peripheral low-pass filtering. Furthermore, durations were expected to increase as filters became stronger or larger; however, as observed previously for varying filter type \cite{Cajar.JVis.2016,Laubrock.JVis.2013} default timing might be adapted with strong or large filters that attenuate most useful spatial frequencies and strongly impede information uptake.

\section{Method}

\subsection{Participants}
Participants were students from the University of Potsdam. For Experiment 1, thirty-two people were tested (10 male, mean age: 21.9 years); for Experiment 2, another thirty-two people, who did not participate in Experiment 1, were tested (11 male, mean age: 22.1 years). They received course credit or fifteen euro for participation in the experiment. All participants had normal or corrected-to-normal vision and normal color discrimination. Participants gave their written informed consent prior to the experiments, which were carried out in accordance with the Code of Ethics of the World Medical Association (Declaration of Helsinki).

\subsection{Apparatus}
Stimuli were presented on a 20$''$ Mitsubishi DiamondPro 2070 CRT monitor with a resolution of 1024 $\times$ 768 pixels and a refresh rate of 120 Hz. A head--chin rest was used to reduce participants' head movements and ensure a constant viewing distance of 60 cm. Gaze position of the right eye was recorded during binocular viewing using an EyeLink 1000 tower mount system (SR Research, Ontario, Canada) with a sampling rate of 1000 Hz. Stimulus presentation was controlled with MATLAB\textsuperscript{\textregistered} (version 2009b; The Mathworks, Natick, MA) using the OpenGL-based Psychophysics Toolbox \cite<PTB3;>{Brainard.SpatVis.1997,Kleiner.Perception.2007,Pelli.SpatVis.1997} and the Eyelink Toolbox \cite{Cornelissen.BehavResMeth.2002}. The highest and lowest possible spatial frequency the monitor could display was 13.2 c/$^\circ$ and 0.03 c/$^\circ$ respectively. System latency from eye movement to screen update (as a sum of frame duration, tracker latency, and eye-velocity computation) was smaller than or equal to 11 ms.

\subsection{Stimuli}
Each experiment consisted of two sessions presenting 240 different color photos of outdoor real-world scenes (120 scenes per session). The same scenes were used in Experiments 1 and 2. Scenes were displayed at a resolution of 1024 $\times$ 768 pixels and a size of 38.7$^\circ$ $\times$ 29.0$^\circ$. They were low-pass filtered in one session and high-pass filtered in the other session. The filtered version of each scene was prepared in advance. Filtering with first-order Butterworth filters was realized in the frequency domain after a Fourier transform of the stimulus, and the filtered image was transformed back into the spatial domain. 

For gaze-contingent filtering in the central or the peripheral visual field, a foreground and a background image were merged in real-time using alpha blending. With peripheral high-pass filtering, for example, the foreground image was the original scene and the background image was the high-pass filtered version of the scene. A 2D hyperbolic tangent with a slope of 0.06 was used as a blending function for creating the alpha mask. The inflection point of the function corresponded to the radius of the gaze-contingent window. The alpha mask was centered at the current gaze position and gave the transparency (i.e., the weighting) of the central foreground image at each point. At the point of fixation, only the foreground image was visible; with increasing eccentricity, the peripheral background image was weighted more strongly until it was fully visible \cite<cf.>{Cajar.JVis.2016}. To avoid jitter due to fixational eye movements, the gaze-contingent display was updated only when the instantaneous eye velocity exceeded 186$^\circ$/s or after a duration of 200 ms.

\subsubsection{Specifics for stimuli in Experiment 1}
In Experiment 1, the gaze-contingent moving window had a constant radius of 3.75$^\circ$. For the three filter levels, filters with cutoff frequencies of 1.26 c/$^\circ$, 3.16 c/$^\circ$, and 7.94 c/$^\circ$ were used; spatial frequencies were attenuated above these thresholds for low-pass filtering and below these thresholds for high-pass filtering. Weak filters preserved a relatively wide band of spatial frequencies and strong filters preserved a relatively narrow band of spatial frequencies; moderate filters were in between. For low-pass filtering, the weak filter had a cutoff frequency of 7.94 c/$^\circ$ and the strong filter had a cutoff frequency of 1.26 c/$^\circ$; for high-pass filtering, the assignment was reversed. All filters were above detection threshold \cite<cf.>{Loschky.VisCogn.2005}. Example stimuli for Experiment 1 are illustrated in Figure \ref{fig:stimuli}a. 

To give the reader a rough estimate how far off into the periphery the cutoff frequencies were visible, we calculated the maximal eccentricities they were visible at according to the cortical magnification principle. Contrast sensitivity for gratings with spatial frequencies measured in cycles/mm of cortex, that is, c/$^\circ$ divided by the cortical magnification factor \emph{M}, is independent of eccentricity \cite{Rovamo.Nature.1978,Rovamo.ExpBrainRes.1979,Virsu.ExpBrainRes.1979}. \citeA[p. 498]{Rovamo.ExpBrainRes.1979} provide four formulas to calculate \emph{M} for different eccentricities---one for each half-meridian of the visual field (superior, inferior, nasal, and temporal), since cortical magnification differs between them. Using these formulas, we estimated the eccentricities at which the cutoff frequencies used in the present experiment became invisible on the superior and temporal half-meridians (along which contrast sensitivity is worst and best respectively). Assuming that gratings with a spatial frequency greater than 3.5 c/mm of cortex are invisible \cite[Fig. 4]{Rovamo.ExpBrainRes.1979}, the cutoff frequencies should become invisible at the following eccentricities: 36.8$^\circ$ and 63.3$^\circ$ for spatial frequencies of 1.26 c/$^\circ$; 18.5$^\circ$ and 28.5$^\circ$ for frequencies of 3.16 c/$^\circ$; 7.9$^\circ$ and 11.6$^\circ$ for frequencies of 7.94 c/$^\circ$.

\subsubsection{Specifics for stimuli in Experiment 2}
\begin{figure}[ht!]	
	\centering
	\includegraphics[width=\textwidth]{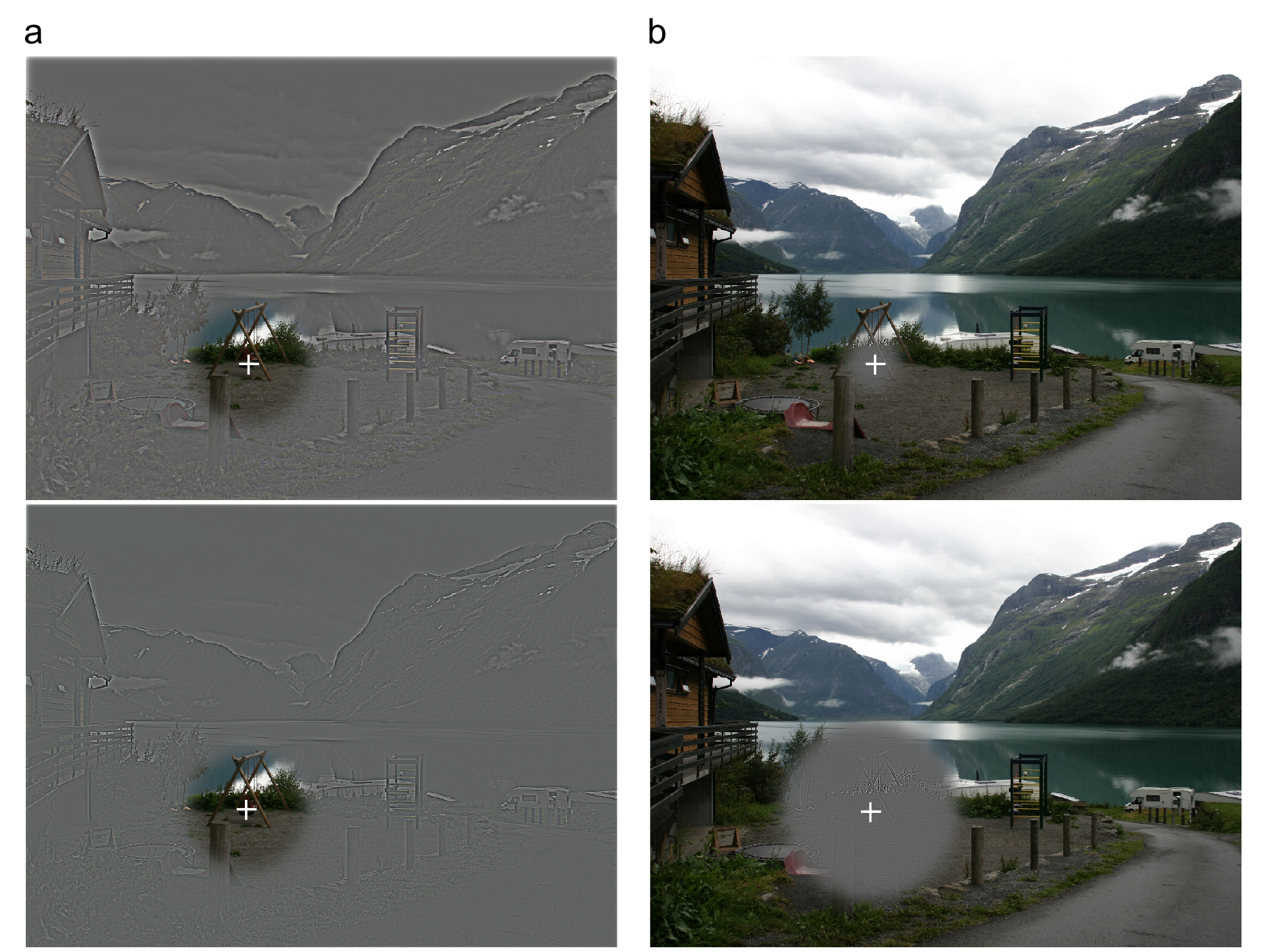}
	\caption{Illustration of four experimental conditions. The white cross illustrates the current gaze position of the viewer. (a) Two filter conditions of Experiment 1: Weak peripheral high-pass filter (top) and moderate peripheral high-pass filter (bottom). (b) Two filter conditions of Experiment 2: Small central high-pass filter (top) and large central high-pass filter (bottom). Note that the illustrated scenes are at a smaller scale than presented in the experiments and therefore do not faithfully reproduce the actual filters used in the experiments.}	
	\label{fig:stimuli} 
\end{figure}

In Experiment 2, filter level was constant with a cutoff frequency of 1.0 c/$^\circ$ for low-pass filtering and 10.0 c/$^\circ$ for high-pass filtering, spatial frequencies at which the parvo and magno cells of the lateral geniculate nucleus are near their peak contrast sensitivity \cite{Derrington.JPhysiol.1984}. According to the aforementioned derivation, these cutoff frequencies approximately become invisible on the superior and temporal half-meridians at the following eccentricities: 42.4$^\circ$ and 75.3$^\circ$ for 1.0 c/$^\circ$; 6.3$^\circ$ and 9.3$^\circ$ for 10.0 c/$^\circ$. The size of the filtered region was varied by varying the radius of the gaze-contingent window, with values of 1.75$^\circ$, 3.75$^\circ$, and 5.75$^\circ$ for the small, medium, and large gaze-contingent window respectively. These window sizes roughly extended to the foveal, the parafoveal, and the peripheral visual field \cite<see>{Larson.JVis.2009}. Thus, small central filters mainly degraded foveal vision whereas large central filters degraded foveal and parafoveal vision; small peripheral filters, on the other hand, degraded only peripheral vision whereas large peripheral filters degraded peripheral and parafoveal vision. Increasing window size from small to large therefore increased the size of the filtered region with central filtering, but decreased the size of the filtered region with peripheral filtering. Example stimuli for Experiment 2 are illustrated in Figure \ref{fig:stimuli}b.

\subsection{Design}
Each experiment consisted of two sessions, one for each filter type (low-pass, high-pass). As each participant completed both sessions, session order was counterbalanced so that half of the participants started with the low-pass session and the other half with the high-pass session. In each session, two filter locations (central visual field, peripheral visual field) were crossed with either three filter levels (weak, moderate, strong) in Experiment 1 or with three filter sizes (small, medium, large) in Experiment 2, yielding six experimental conditions. Two control conditions were added as a lower and upper baseline, presenting scenes either without any filtering (unfiltered control) or entirely filtered (filtered control). For the filtered control condition in Experiment 1, the moderate filter was used. A Latin square design assured counterbalancing of condition--image assignments across participants. Images were presented in random order.

\subsection{Procedure}
For each experiment, data were collected in two one-hour sessions. The eye-tracker was calibrated at the beginning of a session and after every fifteen trials. Each trial started with a fixation point in the center of the screen. The scene was revealed after the point had been fixated for at least 150 ms within two seconds from trial start; otherwise a recalibration was automatically scheduled. 

Each session started with four practice trials to acquaint participants with the gaze-contingent display. Each scene was presented for twelve seconds in one of the eight conditions. Viewers were asked to explore each scene carefully in preparation for a memory question on scene content with three response alternatives presented after the trial. The questions were constructed to be rather difficult to answer, thus encouraging participants to inspect the scenes carefully throughout the entire twelve seconds of scene presentation. There were no questions regarding colors in a scene. Questions typically asked about the presence or absence of certain objects in the scene (e.g., ``Which of the three following objects was not present in the scene?''), about the location of objects (e.g., ``Where was the suitcase standing?''), or about the number of certain objects (e.g., ``How many people were present in the scene?''). The memory question for the scene in Figure \ref{fig:stimuli}, for example, was ``Which object was seen on the playground?'', with the response alternatives ``Sandbox'', ``Trampoline'', or ``Slide'' (correct answer: ``Trampoline'').

\subsection{Data preparation}
Saccades were detected using a velocity-based adaptive algorithm \cite{Engbert.VisionRes.2003,Engbert.PNAS.2006}. A total of 480 trials (6\%) of Experiment 1 and 478 trials (6\%) of Experiment 2 were removed owing to poor recording or too much data loss. Single fixations and saccades were removed if they neighbored eye blinks or were outside of the monitor area. The first and the last event of a trial were also excluded from analyses, since they were associated with scene onset and offset. Glissades following a saccade were assigned to the saccade; if more than one glissade followed a saccade, the glissades and their adjacent fixation and saccade were removed \cite<cf.>{Cajar.JVis.2016,Laubrock.JVis.2013}. In total, 273,716 fixations and 283,692 saccades were left for analyses of Experiment 1; 256,711 fixations and 266,912 saccades were left for analyses of Experiment 2.

\subsection{Data analyses}
For each experiment, data from both sessions were merged for the analyses. Fixation durations and saccade amplitudes were analyzed using linear mixed-effects models (LMM), and task performance was analyzed using binomial generalized linear mixed-effects models (GLMM) with a logit link function as implemented in the \emph{lme4} package \cite{Bates.lme4.2015}; this package is supplied in the \emph{R} system for statistical computing \cite<version 3.2.3;>{RCoreTeam.2015}. In addition to fixed effects for the experimental manipulations, (G)LMMs also account for random effects (i.e., variance components) due to differences between participants and scenes, which reduces unexplained variance. Fixed-effects parameters were estimated via user-defined contrasts testing for eight main effects and seven interaction effects for the three experimental factors (filter type, filter location, filter level/size) and the two control conditions.

All GLMM analyses provide regression coefficients, standard errors, \emph{z}-values, and \emph{p}-values for fixed effects. LMM analyses only provide regression coefficients, standard errors, and \emph{t}-values, because the degrees of freedom are not known exactly for LMMs. For large data sets as in the present experiments, however, the \emph{t}-distribution has converged to the standard normal distribution for all practical purposes \cite[Note 1]{Baayen.JMemLang.2008}. Therefore, \emph{t}-statistics exceeding an absolute value of 1.96 were considered statistically significant on the two-tailed 5\% level.
 
Because distributions of fixation durations and saccade amplitudes were positively skewed, both variables were transformed before model fitting to approximate normally distributed model residuals. To find a suitable transformation, the optimal $\lambda$-coefficient for the Box-Cox power transformation \cite{Box.JRoyalStatSoc.1964} was estimated with the \emph{boxcox} function of the \emph{MASS} package \cite{Venables.MASS.2002}, with $y(\lambda) = (y^\lambda-1)/\lambda$, if $\lambda \neq 0$ and $log(y)$, if $\lambda = 0$ \cite<cf.>{Cajar.JVis.2016}. For fixation durations, $\lambda$ was near zero for both experiments ($\lambda$= 0.02 and $\lambda$= 0.06 for Experiments 1 and 2 respectively), so the log-transformation was chosen; for saccade amplitudes, $\lambda$ = 0.22 (Experiment 1) and $\lambda$ = 0.18 (Experiment 2) was used.

Raw data and analyses for both experiments are available at the Potsdam Mind Research Repository (PMR2) at http://read.psych.uni-potsdam.de/index.php?option=com\_content\&view=article\&id=149.

\section{Results of Experiment 1}

\subsection{Fixation durations}
\begin{figure}[ht!]	
	\centering
	\includegraphics[width=\textwidth]{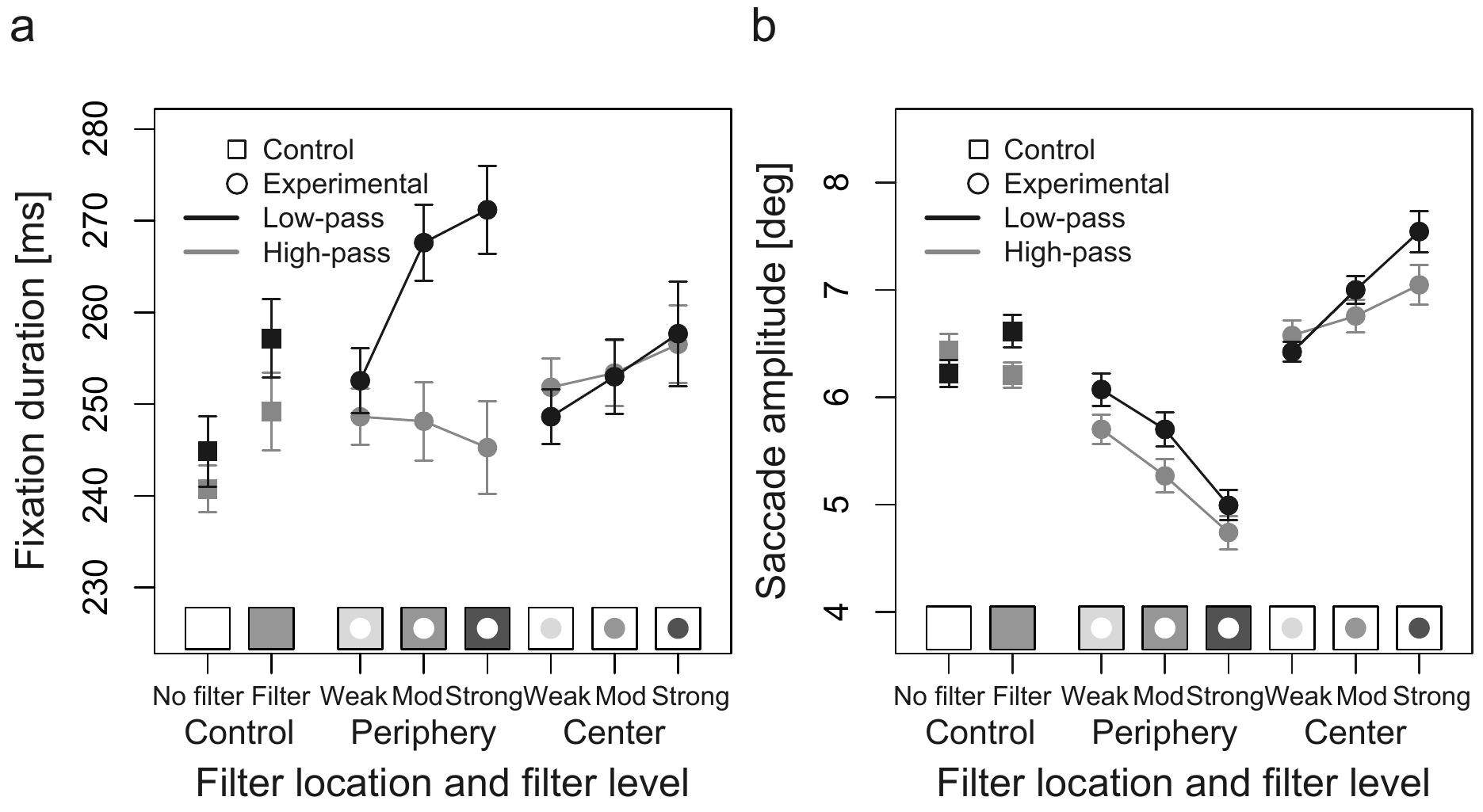}
	\caption{Eye-movement behavior in Experiment 1. (a) Mean fixation durations. (b) Mean saccade amplitudes. Error bars are 95\% within-subject confidence intervals. The boxes at the bottom of the plots illustrate the different filter conditions---the gray area indicates the filtered part and the white area indicates the unfiltered part of the visual field; the gray level from light to dark indicates the level of filtering from weak to strong.}
	\label{fig:eyemovements_Exp1}
\end{figure}

Mean fixation durations for all conditions are illustrated in Figure \ref{fig:eyemovements_Exp1}a. Consistent with the notion that fixation durations adapt to visual-cognitive processing difficulty, mean fixation durations increased when the scene was partly or entirely filtered, $b = 4.37 \cdot 10^{-2}, \mathit{SE} = 2.36 \cdot 10^{-3}, t = 18.5$. Interestingly, on average, filtering only the central or the peripheral part of the visual field increased mean fixation durations more than filtering of the entire scene did ($b = 1.12 \cdot 10^{-2}, \mathit{SE} = 2.44 \cdot 10^{-3}, t = 4.6$), an effect mainly driven by the strong increase of fixation durations with moderate and strong peripheral low-pass filters compared with the filtered controls (see Figure \ref{fig:eyemovements_Exp1}a). Comparing the filtered controls showed longer fixation durations with low-pass than with high-pass filtered scenes ($b = 2.35 \cdot 10^{-2}, \mathit{SE} = 5.51 \cdot 10^{-3}, t = 4.3$). 

Regarding gaze-contingent central and peripheral filtering, significant main effects of filter type ($b = 2.29 \cdot 10^{-2}, \mathit{SE} = 3.64 \cdot 10^{-3}, t = 6.3$) and filter location ($b = 1.56 \cdot 10^{-2}, \mathit{SE} = 1.84 \cdot 10^{-3}, t = 8.5$) indicated longer fixation durations with low-pass than with high-pass filtering and with peripheral than with central filtering. These main effects, however, were qualified by a strong interaction of filter type and filter location, $b = 2.79 \cdot 10^{-2}, \mathit{SE} = 1.84 \cdot 10^{-3}, t = 15.2$. First, filter type had almost no effect with central filtering, but a strong effect with peripheral filtering where fixation durations increased markedly with low-pass, but not with high-pass filtering. Second, central filtering provoked longer fixations than peripheral high-pass filtering, but shorter fixations than peripheral low-pass filtering. 

Main effects of filter level indicated longer fixation durations with stronger filters---fixations were longer with moderate and strong filters than with weak filters ($b = 2.20 \cdot 10^{-2}, \mathit{SE} = 1.94 \cdot 10^{-3}, t = 11.3$) and longer with strong filters than with moderate filters ($b = 9.01 \cdot 10^{-3}, \mathit{SE} = 2.26 \cdot 10^{-3}, t = 4.0$). However, these effects of filter level were only pronounced with low-pass filtering, as indicated by interactions of filter level and filter type ($b = 1.83 \cdot 10^{-2}, \mathit{SE} = 1.94 \cdot 10^{-3}, t = 9.4$ for weak versus moderate and strong filters, and $b = 8.37 \cdot 10^{-3}, \mathit{SE} = 2.26 \cdot 10^{-3}, t = 3.7$ for moderate versus strong filters). Interestingly, filter level hardly affected fixation durations with high-pass filtering, regardless of filter location. Filter level had the biggest effect with peripheral low-pass filtering, where mean fixation durations increased markedly from weak to moderate and strong filters. Therefore, an interaction of filter location and filter level indicated a stronger increase of fixation durations from weak to moderate and strong filters with peripheral filtering than with central filtering, $b = 8.75 \cdot 10^{-3}, \mathit{SE} = 1.94 \cdot 10^{-3}, t = 4.5$. All variance components and fixed effects of the LMM can be found in the Appendix (Table \ref{tab:LMM_FD_Exp1}).

\subsection{Saccade amplitudes}
Mean saccade amplitudes for all conditions are illustrated in Figure \ref{fig:eyemovements_Exp1}b. Effects turned out as expected. There was a strong main effect of filter location ($b = 8.14 \cdot 10^{-2}, \mathit{SE} = 1.11 \cdot 10^{-3}, t = 73.54$), as mean saccade amplitudes were longer with central than with peripheral filtering. More precisely, central filtering elicited longer saccades than the control conditions, whereas peripheral filtering elicited shorter saccades than the control conditions. This indicates a preference for unfiltered scene regions as saccade targets. Due to this opposing viewing behavior with central and peripheral filtering, there was no main effect of filter level, but strong interactions of filter level and filter location ($b = -3.67 \cdot 10^{-2}, \mathit{SE} = 1.17 \cdot 10^{-3}, t = -31.33$ for weak versus moderate and strong filters, and $b = -2.65 \cdot 10^{-2}, \mathit{SE} = 1.36 \cdot 10^{-3}, t = -19.48$ for moderate versus strong filters)---as the filter became stronger, amplitudes increasingly lengthened with central filtering, but increasingly shortened with peripheral filtering. Thus, saccade target selection deviated more and more from normal as processing difficulty increased. Accordingly, low-pass filtering led to longer saccades than high-pass filtering with both filter locations ($b = 1.31 \cdot 10^{-2}, \mathit{SE} = 3.28 \cdot 10^{-3}, t = 3.98$)---because low spatial frequencies are more useful than high frequencies for peripheral target selection, more long saccades occurred with peripheral low-pass filtering; however, low spatial frequencies are less useful than high frequencies for detailed foveal analysis, so that fewer short saccades occurred with central low-pass filtering. The difference between low-pass and high-pass filters was more pronounced with peripheral than with central filtering, leading to an interaction of filter type and filter location ($b = 1.01 \cdot 10^{-2}, \mathit{SE} = 1.11 \cdot 10^{-3}, t = 9.13$). With peripheral filtering, filter level had a similar effect on saccade amplitudes with low-pass and high-pass filtering; with central filtering, on the other hand, filter level had a stronger effect with low-pass than with high-pass filtering. This is reflected by three-way interactions between filter type, filter location and filter level, with $b = -1.01 \cdot 10^{-2}, \mathit{SE} = 1.17 \cdot 10^{-3}, t = -8.62$ for weak versus moderate and strong filters, and $b = -7.41 \cdot 10^{-3}, \mathit{SE} = 1.36 \cdot 10^{-3}, t = -5.45$ for moderate versus strong filters. Saccade amplitudes were least affected by weak low-pass or high-pass filters, showing that these filters preserved a sufficient amount of high or low spatial frequencies respectively for near-normal spatial viewing behavior. 

Saccade amplitudes in the filtered control conditions were similar to amplitudes in the unfiltered control condition, with low-pass filtered scenes provoking slightly longer saccades than their unfiltered control and high-pass filtered scenes provoking slightly shorter saccades than their unfiltered control (see Figure \ref{fig:eyemovements_Exp1}b). Compatible with the results for central and peripheral filtering, completely low-pass filtered scenes led to longer saccades than high-pass filtered scenes, $b = 2.36 \cdot 10^{-2}, \mathit{SE} = 4.12 \cdot 10^{-3}, t = 5.73$. All variance components and fixed effects of the LMM can be found in the Appendix (Table \ref{tab:LMM_SA_Exp1}).
 
\subsection{Task performance}
\begin{figure}[ht!]	
	\centering
	\includegraphics[width=\textwidth]{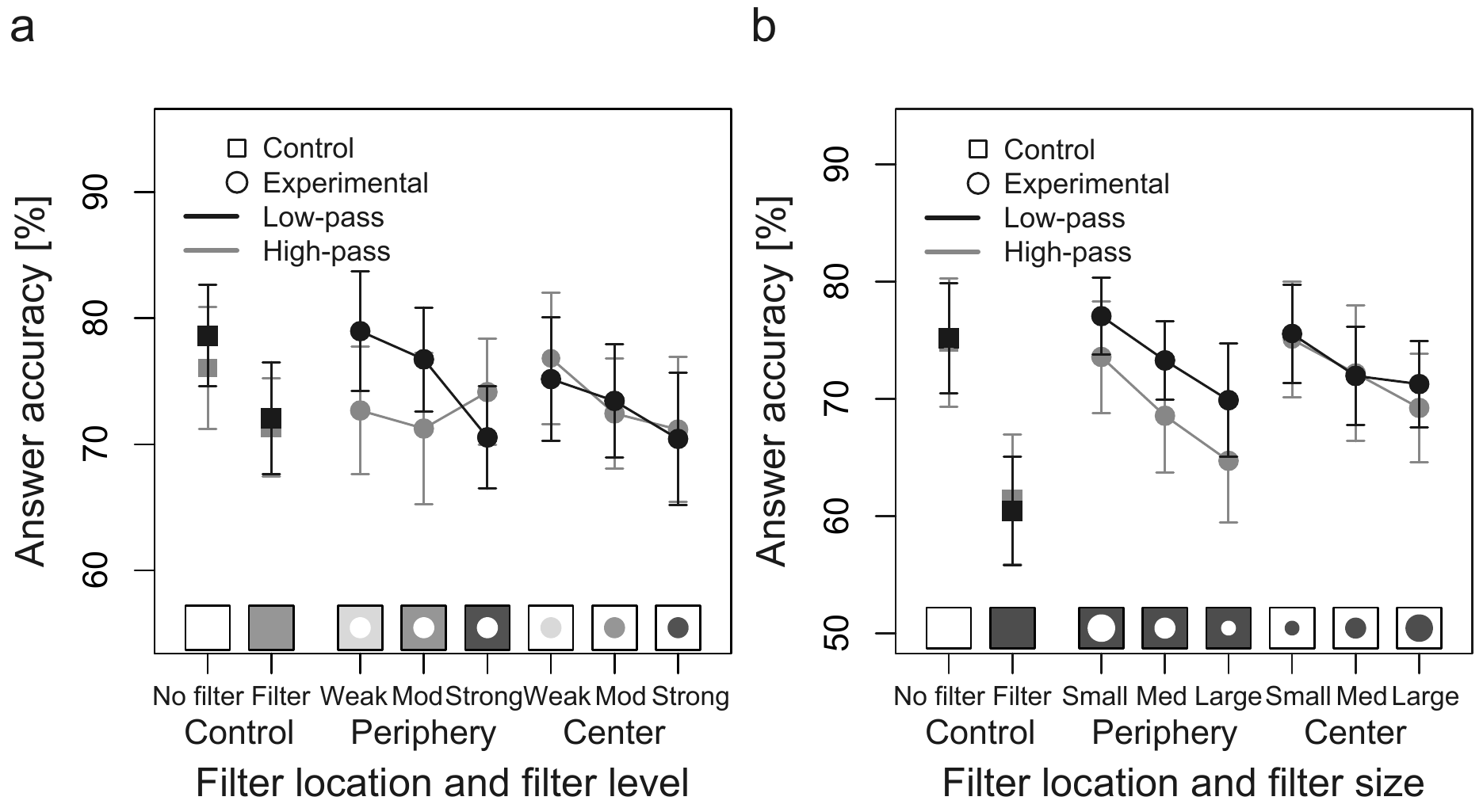}
	\caption{Mean proportions of correct answers to the memory questions in Experiment 1 (a) and Experiment 2 (b). Error bars are 95\% within-subject confidence intervals. The boxes at the bottom of the plots illustrate the different filter conditions---the gray area indicates the filtered part and the white area indicates the unfiltered part of the visual field. The gray level from light to dark in (a) indicates the level of filtering from weak to strong.}
	\label{fig:taskperformances}
\end{figure}

Mean proportion of correct answers to the memory questions are illustrated in Figure \ref{fig:taskperformances}a. Task performance decreased when the scene was partly or entirely filtered, $b = -2.55 \cdot 10^{-1}, \mathit{SE} = 9.37 \cdot 10^{-2}, z = -2.72, p = .006$. Performance decreased with moderate and strong filters compared to weak filters, $b = -2.11 \cdot 10^{-1}, \mathit{SE} = 7.42 \cdot 10^{-2}, z = -2.84, p = .004$. An interaction between filter type and moderate versus strong filters ($b= 1.74 \cdot 10^{-1}, \mathit{SE} = 8.36 \cdot 10^{-2}, z = 2.08, p = .038$) occurred because strong filters interfered more with task performance than moderate filters with low-pass filtering, but not with high-pass filtering. There were no other significant effects.

\subsection{Summary} 
In most conditions eye-movement behavior was increasingly modulated by increasing processing difficulty due to spatial-frequency filtering. Task performance decreased when scenes were filtered and decreased more as filters became stronger. Fixation durations increased with filtering. They were markedly longer with peripheral low-pass filtering than with peripheral high-pass filtering, suggesting that more processing time was invested when the degraded information could still be used efficiently for gaze control. Increasing filter level prolonged fixation durations with low-pass filtering, but had almost no effect with high-pass filtering, which we interpret as default timing. Instead, viewing behavior was adapted to stronger high-pass filters by modulation of saccade amplitudes.

As expected, saccades lengthened with central filtering and shortened with peripheral filtering compared to the unfiltered control condition, indicating a preference for unfiltered scene regions as saccade targets. Effects scaled with filter type and filter level. First, saccade amplitudes deviated more from normal viewing behavior when the available spatial frequencies were less useful for saccade target selection or foveal analysis (i.e., with peripheral high-pass filtering and central low-pass filtering respectively). Second, amplitudes monotonically lengthened with central filtering and shortened with peripheral filtering as filter level increased. Thus, the stronger the filter, the greater the bias to avoid the filtered region. 

\section{Results of Experiment 2}

\subsection{Fixation durations}
\begin{figure}[htbp]	
	\centering
	\includegraphics[width=\textwidth]{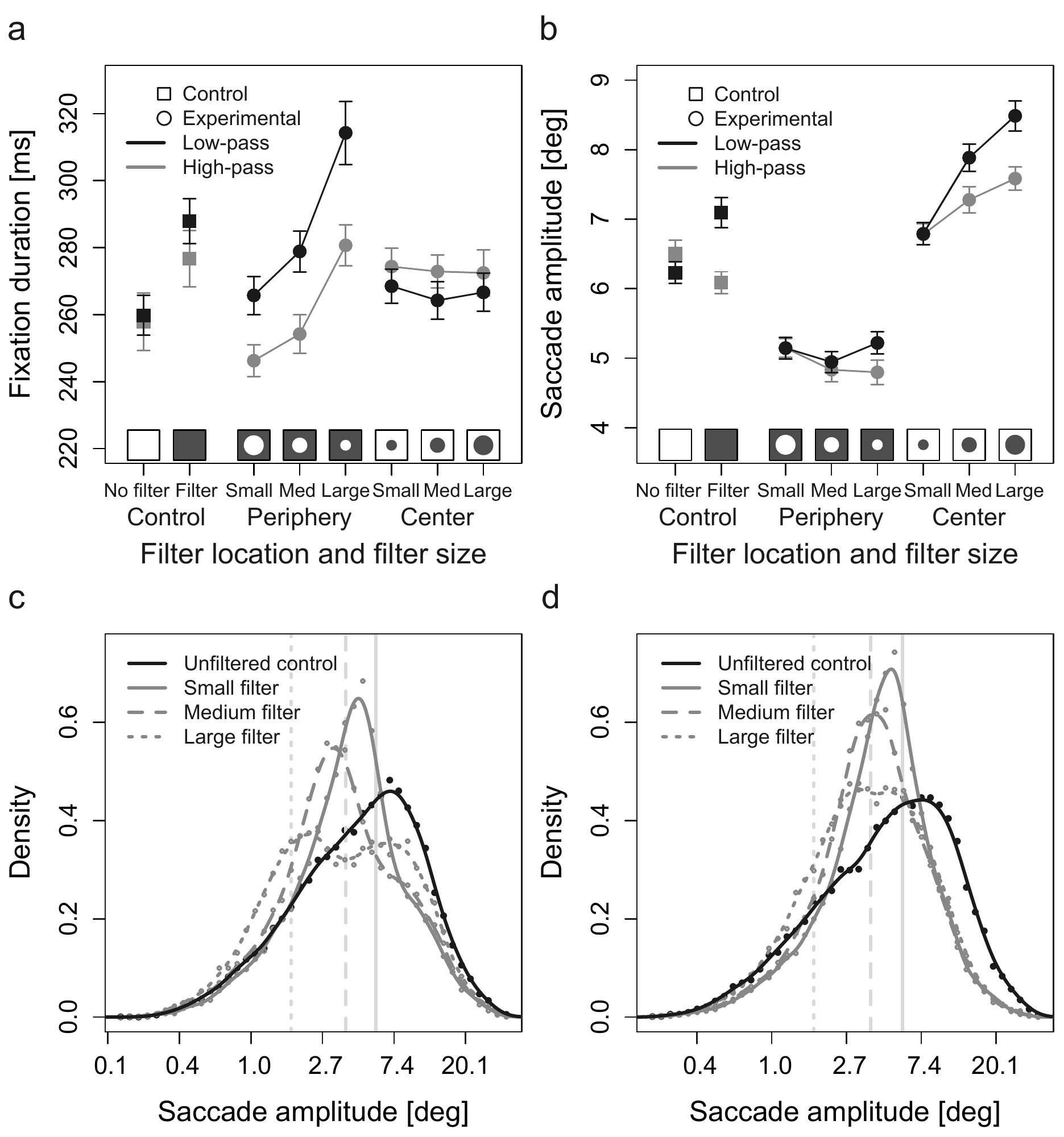}
	\caption{Eye-movement behavior in Experiment 2. (a-b) Mean fixation durations (a) and mean saccade amplitudes (b). Error bars are 95\% within-subject confidence intervals. The boxes at the bottom of the plots illustrate the different filter conditions---the dark area indicates the filtered part and the white area indicates the unfiltered part of the visual field. (c-d) Distributions of saccade amplitudes for peripheral low-pass filtering (c) and peripheral high-pass filtering (d). Note that the abscissa is log-scaled. Lines represent kernel density estimates using a Gaussian kernel as implemented in the R function \emph{density}. The area under each curve adds up to one. Vertical gray lines represent the gaze-contingent window radii of 1.75$^\circ$ (dotted line), 3.75$^\circ$ (dashed line), and 5.75$^\circ$ (solid line) corresponding to large, medium, and small peripheral filters respectively.}
	\label{fig:eyemovements_Exp2}
\end{figure}

Mean fixation durations for all conditions are illustrated in Figure \ref{fig:eyemovements_Exp2}a. Fixation durations increased when the scene was partly or entirely filtered compared with the unfiltered control conditions, $b = 5.41 \cdot 10^{-2}, \mathit{SE} = 2.49 \cdot 10^{-3}, t = 21.76$. Comparing the filtered controls revealed longer fixation durations with low-pass filtering than with high-pass filtering, $b = 1.91 \cdot 10^{-2}, \mathit{SE} = 5.78 \cdot 10^{-3}, t = 3.31$. In contrast to Experiment 1, completely filtered scenes provoked longer mean fixation durations than central or peripheral filtering did ($b = 3.21 \cdot 10^{-2}, \mathit{SE} = 2.66 \cdot 10^{-3}, t = 12.09$), but fixation durations were again longer with peripheral than with complete low-pass filtering when the filter was large (see Figure \ref{fig:eyemovements_Exp2}a).
 
With central and peripheral spatial-frequency filtering the LMM revealed similar effects as in Experiment 1. Main effects of filter type ($b = 2.99 \cdot 10^{-2}, \mathit{SE} = 3.54 \cdot 10^{-3}, t = 8.44$) and filter location ($b = 1.68 \cdot 10^{-2}, \mathit{SE} = 1.92 \cdot 10^{-3}, t = 8.74$) emerged, indicating longer mean fixation durations with low-pass filtering than with high-pass filtering and longer mean fixation durations with peripheral filtering than with central filtering. These main effects were qualified by a strong interaction of filter type and filter location ($b = 5.37 \cdot 10^{-2}, \mathit{SE} = 1.92 \cdot 10^{-3}, t = 27.96$), as fixation durations were longer with central high-pass and peripheral low-pass filtering than with central low-pass and peripheral high-pass filtering. Consistent with Experiment 1, the difference between low-pass and high-pass filtering was larger with peripheral than with central filtering. 

Most remarkable was the effect of filter size. Two main effects indicated that fixation durations increased as filters became larger ($b = 3.77 \cdot 10^{-2}, \mathit{SE} = 2.02 \cdot 10^{-3}, t = 18.64$ for small versus medium and large filters, and $b = 4.69 \cdot 10^{-2}, \mathit{SE} = 2.37 \cdot 10^{-3}, t = 19.77$ for medium versus large filters). Interestingly, this increase occurred only with peripheral filtering, particularly from medium to large filters---with central filtering, filter size had no effect on fixation durations. This is reflected in two strong interactions of filter location and filter size ($b = 4.66 \cdot 10^{-2}, \mathit{SE} = 2.02 \cdot 10^{-3}, t = 23.05$ for small versus medium and large filters, and $b = 4.42 \cdot 10^{-2}, \mathit{SE} = 2.37 \cdot 10^{-3}, t = 18.66$ for medium and large filters). Thus, although fixation durations increased with spatial-frequency filtering in the foveal region (small central filter) compared with the unfiltered control, a further extension of the filter to the parafovea did not increase fixation durations any further.

Figure \ref{fig:eyemovements_Exp2}a shows that, similar to Experiment 1, fixation durations were closest to the unfiltered control condition with central low-pass filtering, peripheral high-pass filtering, and peripheral low-pass filtering with the smallest filter. With small and medium peripheral high-pass filters mean fixation durations even dropped below the mean of the unfiltered control condition. All variance components and fixed effects of the LMM can be found in the Appendix (Table \ref{tab:LMM_FD_Exp2}).

\subsection{Saccade amplitudes}
Mean saccade amplitudes for all conditions are illustrated in Figure \ref{fig:eyemovements_Exp2}b. Results largely turned out as in Experiment 1 and showed adaptation of saccade amplitudes to increasing processing difficulty. A strong main effect of filter location occurred ($b = 9.89 \cdot 10^{-2}, \mathit{SE} = 8.53 \cdot 10^{-4}, t = 115.88$), because saccade amplitudes increased with central filtering and decreased with peripheral filtering. A main effect of filter type ($b = 1.17 \cdot 10^{-2}, \mathit{SE} = 2.34 \cdot 10^{-3}, t = 5.00$) reflected longer saccades with low-pass filtering than with high-pass filtering. As indicated by an interaction of filter type and filter location ($b = 2.01 \cdot 10^{-2}, \mathit{SE} = 8.53 \cdot 10^{-4}, t = 23.49$), this difference between filter types was stronger with central filtering than with peripheral filtering. Saccade amplitudes were similar with peripheral low-pass and high-pass filters of small or medium size. As in Experiment 1, completely low-pass filtered scenes provoked longer saccade amplitudes than completely high-pass filtered scenes ($b = 4.51 \cdot 10^{-2}, \mathit{SE} = 3.10 \cdot 10^{-3}, t = 14.56$). 

Main effects of filter size indicated increasing mean saccade amplitudes as filters became larger, that is, longer amplitudes with medium and large filters than with small filters ($b = 1.02 \cdot 10^{-2}, \mathit{SE} = 8.98 \cdot 10^{-4}, t = 11.37$), and longer amplitudes with medium than with large filters ($b = 4.37 \cdot 10^{-3}, \mathit{SE} = 1.05 \cdot 10^{-3}, t = 4.15$). However, these main effects were largely due to the effects of central filtering on saccade amplitudes and were therefore qualified by interactions of filter size and filter location ($b = 3.15 \cdot 10^{-2}, \mathit{SE} = 8.98 \cdot 10^{-4}, t = 35.09$ for small versus medium and large filters, and $b = 9.51 \cdot 10^{-3}, \mathit{SE} = 1.05 \cdot 10^{-3}, t = 9.03$ for medium versus large filters). The increase of saccade amplitudes with increasing central filter size was more pronounced with low-pass filters than with high-pass filters, as indicated by interactions of filter type and filter size ($b = 1.02 \cdot 10^{-2}, \mathit{SE} = 8.98 \cdot 10^{-4}, t = 11.32$ for small versus medium and large filters, and $b = 6.57 \cdot 10^{-3}, \mathit{SE} = 1.05 \cdot 10^{-3}, t = 6.24$ for medium versus large filters). 

Interestingly, mean saccade amplitudes were hardly affected by filter size with peripheral filtering, which is in contrast to previous studies demonstrating a decrease of mean amplitudes with increasing low-pass filter size \cite{Nuthmann.VisCogn.2013,Nuthmann.JExpPsycholHuman.2014}. Inspection of the distributions of saccade amplitudes for peripheral low-pass and high-pass filtering (Figures \ref{fig:eyemovements_Exp2}c and \ref{fig:eyemovements_Exp2}d respectively) sheds light on the absent effect of filter size. In the filter conditions, the modes of the distributions mostly shifted toward the respective radius of the gaze-contingent window, and the amount of short saccades increased with increasing filter size. Thus, viewers programmed more saccades inside or to the border of the unfiltered central region with increasing filter size. With low-pass filtering, this effect was counteracted by an increased number of long saccades from small to large filters, causing the effect of decreasing mean saccade amplitudes with increasing filter size to disappear. The effects might reflect a viewing strategy where smaller parts of the scene are inspected through a series of short saccades inside the unfiltered central region, followed by long saccades to target new parts of the scene. All variance components and fixed effects of the LMM can be found in the Appendix (Table \ref{tab:LMM_SA_Exp2}).

\subsection{Task performance}
Mean proportion of correct answers to the memory questions are illustrated in Figure \ref{fig:taskperformances}b. Compared with unfiltered scene viewing, task performance decreased when the scene was partly or entirely filtered ($b = -2.89 \cdot 10^{-1}, \mathit{SE} = 9.06 \cdot 10^{-2}, z = -3.19, p = .001$), with the decrease being stronger when the entire scene was filtered ($b = -6.33 \cdot 10^{-1}, \mathit{SE} = 8.57 \cdot 10^{-2}, z = -7.38, p < .001$). Furthermore, task performance decreased with increasing filter size, as shown by two main effects of filter size ($b = -2.87 \cdot 10^{-1}, \mathit{SE} = 7.24 \cdot 10^{-2}, z = -3.96, p < .001$ for small versus medium and large filters, and $b = -1.78 \cdot 10^{-1}, \mathit{SE} = 8.13 \cdot 10^{-2}, z = -2.19, p < .029$ for medium versus large filters). There were no other significant effects.

\subsection{Summary}
Spatial-frequency filtering with varying filter size clearly modulated eye-movement behavior. Task performance decreased when the scene was filtered and decreased more when filters became larger. Fixation durations increased in most filter conditions. Central high-pass and peripheral low-pass filtering elicited longer fixation durations than central low-pass and peripheral high-pass filtering, suggesting that more processing time was invested when the available spatial frequencies matched the viewers' tasks (fovea: analysis of details, periphery: saccade target selection). Increasing filter size prolonged fixation durations monotonically with peripheral filtering, but had no effect on fixation durations with central filtering. The latter implies that fixation durations were mainly controlled by processing difficulty in the fovea. 

Saccade amplitudes largely adapted to increasing processing difficulty. Amplitudes increased with central filtering and decreased with peripheral filtering. With medium and large filters, amplitudes deviated more from unfiltered viewing when spatial frequencies were potentially less useful for processing. Furthermore, increasing filter size provoked more long saccades with central filtering and more short saccades with peripheral filtering.
 
Effects of filter size were pronounced with either mean fixation durations or saccade amplitudes, but not both. With peripheral filtering, filter size had clear effects on mean fixation durations, but not on mean saccade amplitudes; with central filtering, the pattern was reversed. This suggests a trade-off between temporal and spatial aspects of eye-movement behavior. 

\section{Discussion}
The present study investigated how the availability of different spatial frequencies in the central or the peripheral visual field affects eye-movement behavior during real-world scene viewing. For this purpose, high-pass or low-pass filters were applied either inside or outside a gaze-contingent window while the respective other region remained unfiltered. Filter level was varied in Experiment 1, and filter size was varied in Experiment 2. Results demonstrate that temporal as well as spatial aspects of eye-movement behavior are modulated by the type, size, and level of filtering.

\subsection{Fixation durations do not always increase with processing difficulty}
Of particular interest was the question how fixation durations adapt to varying processing difficulty due to spatial-frequency filtering. Generally, fixation durations are assumed to increase as visual-cognitive processing difficulty increases. Accordingly, prior research reported increased fixation durations when low spatial frequencies are filtered from the central or the peripheral visual field during scene viewing \cite{Loschky.JExpPsycholAppl.2002,Loschky.VisCogn.2005,Nuthmann.VisCogn.2013,Nuthmann.JExpPsycholHuman.2014,Parkhurst.INBOOK.2000,vanDiepen.Perception.1998}, with the effects getting stronger with increasing filter size \cite{Loschky.JExpPsycholAppl.2002,Nuthmann.VisCogn.2013,Nuthmann.JExpPsycholHuman.2014,Parkhurst.INBOOK.2000}. 

In contrast to these findings, the present results indicate that fixation durations increase with higher processing difficulty due to spatial-frequency filtering only when difficulty is moderately increased. The increase of fixation durations was more pronounced with central high-pass and peripheral low-pass filtering than with central low-pass and peripheral high-pass filtering. This result replicates our findings from previous studies \cite{Cajar.JVis.2016,Laubrock.JVis.2013} and suggests that viewers invest more time for processing spatial frequencies that are potentially more useful for foveal analysis of details (high spatial frequencies) and peripheral target selection (low spatial frequencies). When processing becomes too difficult and information cannot be efficiently used for gaze control, fixation durations do not prolong any further. In this sense, viewing behavior appears to be quite economical. 

Most impressively, this behavior is reflected in the effects of filter level and filter size. First, fixation durations were unaffected by increasing filter level with peripheral high-pass filtering. High spatial frequencies cannot be resolved in the low-resolution periphery and are therefore hardly useful for saccade target selection. Thus, increasing fixation durations with increasing level of high-pass filtering was probably not worthwhile. Increasing the level of peripheral low-pass filtering, on the other hand, prolonged fixation durations, indicating that strong low-pass filtered information can still be used for peripheral target selection. This result contradicts Loschky and colleagues \cite{Loschky.JExpPsycholAppl.2002,Loschky.VisCogn.2005}, who found that the level of detectable peripheral low-pass filtering hardly affected fixation durations. Second, fixation durations were unaffected by increasing filter size with both central high-pass and low-pass filtering. Compared with unfiltered scene viewing, fixation durations increased with spatial-frequency filtering in the foveal visual field, but additional filtering of the parafovea did not prolong fixation durations any further \cite<contrary to>{Nuthmann.JExpPsycholHuman.2014}. Thus, fixation durations appeared to be mainly controlled by processing difficulty in the fovea. Increasing filter size in the peripheral visual field, though, increased fixation durations monotonically \cite<replicating>{Loschky.JExpPsycholAppl.2002,Nuthmann.VisCogn.2013,Nuthmann.JExpPsycholHuman.2014,Parkhurst.INBOOK.2000}. Still, fixation durations were similar to or shorter than the unfiltered control condition with medium and large peripheral high-pass filters respectively. Presumably, foveal and parafoveal scene information could be processed rather efficiently in these conditions, but little time was invested for saccade target selection in an apparently uniform gray periphery where the extraction of scene information was rather difficult. 

We conclude that fixation durations prolong with increasing processing difficulty due to the type, level, and size of spatial frequency filters as long as the available information can be analyzed in a reasonable amount of time and is critical for gaze control; otherwise, stimulus-independent timing takes over.  
	
\subsection{Peripheral information is critical for the control of fixation duration}
Interestingly, fixation durations were affected more strongly by increasing processing difficulty in the peripheral visual field than in the central visual field. First, mean differences in fixation durations between filter types were notably larger with peripheral filtering in both experiments, suggesting that the kind of available spatial-frequency information is more important in the peripheral than in the central visual field. This seems reasonable---since the fovea and parafovea are sensitive to a larger band of spatial frequencies than the periphery, they cope better with missing frequencies when processing the scene. 

Second, foveal analysis was less affected by increasing processing difficulty due to increasing filter size and filter level than peripheral selection. With both low-pass and high-pass filtering, fixation durations increased with central filtering, but were not affected by the size of the central filter and increased little as filter level became stronger. Based on these results, we conclude that (i) the fovea is critical for the control of fixation durations whereas the parafovea plays a subordinate role, and (ii) the level of scene degradation via spatial filtering of the foveal stimulus has little effect on fixation durations. With peripheral filtering, however, filter level and filter size had considerable effects on fixation durations. Mean durations increased with both filter types as the filtered part of the scene became larger, thus making saccade target selection increasingly difficult. With peripheral low-pass filtering, fixation durations also increased with increasing level of filtering.

The present findings suggest that peripheral vision is more sensitive to the kind of available spatial-frequency information than central vision. Contrary to the common notion that fixation durations are dominated by foveal processing, our results show that fixation durations can be controlled considerably by peripheral processing.

\subsection{Switching costs between filtered and unfiltered scene regions}
Peripheral low-pass filtering with moderate and strong filters (Experiment 1) and with large filters (Experiment 2) provoked mean fixation durations that were longer than the mean duration in the filtered control condition in which the entire scene was low-pass filtered. Thus, gaze-contingent filtering entailed switching costs between the filtered peripheral and the unfiltered central scene region when the filter was strong or large enough. \citeA{Reingold.BehavResMethInstr.2002} found similar effects, with longer saccade latencies to a peripheral target when only the peripheral visual field was low-pass filtered compared to low-pass filtering of the entire scene. The authors reasoned that low-pass filtering of only part of the visual field increases the saliency of the unfiltered region; as a consequence, the competition for attention between the filtered and unfiltered scene region might increase and therefore prolong fixation durations compared to viewing entirely filtered scenes. 

\subsection{Saccade amplitudes adapt to processing difficulty}
Saccadic behavior strongly adapted to the manipulated parameters of the stimulus. Compared to viewing unfiltered scenes, mean saccade amplitudes lengthened with central filtering and shortened with peripheral filtering. This effect replicates previous findings \cite{Cajar.JVis.2016,Foulsham.AttenPerceptPsychophys.2011,Laubrock.JVis.2013,Loschky.JExpPsycholAppl.2002,Nuthmann.VisCogn.2013,Nuthmann.JExpPsycholHuman.2014,Reingold.BehavResMethInstr.2002,vanDiepen.Perception.1998} and indicates a preference for unfiltered scene regions as saccade targets---central filtering elicits a bias for targeting scene regions farther away from the current fixation position, whereas peripheral filtering elicits more focussed gaze patterns in the unfiltered central region. This opposing viewing behavior with central and peripheral filtering strengthened as scene processing became more difficult. First, saccades increasingly lengthened with central filtering and increasingly shortened with peripheral filtering as filter level and filter size increased. Second, saccade amplitudes increased more when the central visual field was low-pass filtered than when it was high-pass filtered, whereas they decreased more when the peripheral visual field was high-pass filtered than when it was low-pass filtered. 

In summary, saccade target selection was progressively impaired by increasing processing difficulty due to filtering. As shown by \citeA{Cajar.JVis.2016}, these modulations of saccadic selection reflect modulations of attention, with smaller saccades indicating a withdrawal of attention from the periphery (i.e., tunnel vision) and larger saccades indicating an attentional bias toward the periphery.

\subsection{Trade-off between fixation duration and saccade amplitude}
Increasing filter size in the peripheral visual field increased mean fixation durations but hardly affected mean saccade amplitudes; increasing filter size in the central visual field, on the other hand, increased saccade amplitudes but left fixation durations unaffected. Varying filter size therefore provoked a viewing behavior with a trade-off between fixation durations and saccade amplitudes \cite<see>{Jacobs.PerceptPsychophys.1986}, with opposing effects for central and peripheral filtering. 

This trade-off was also evident with increasing processing difficulty due to filter level---as filter level increased with peripheral high-pass filtering, fixation durations were unaffected, but saccade amplitudes increasingly shortened. Furthermore, mean fixation durations were similar to or even below the mean of the unfiltered control condition with medium and large peripheral high-pass filters, whereas saccade amplitudes shortened. We provide the following explanation for these effects. High-pass filtering inherently attenuates luminance, contrast, and color information of the stimulus. Therefore, the segregation of peripheral objects from their background becomes harder with increasing filter level; with a strong high-pass filter, the peripheral scene resembles a uniform gray background. Thus, selecting saccade targets from strongly high-pass filtered peripheral scene regions was probably too difficult, causing saccade targets to be chosen either from the unfiltered central part of the scene (hence shorter saccade amplitudes) or less carefully from the filtered periphery. Both strategies can reduce the time needed to select the next saccade target and thus shorten mean fixation durations as observed. However, when the peripheral filter gets so large that it approaches foveal vision, the analysis of both the peripheral and the central stimulus is affected, which might explain the observed increase of mean fixation duration with large peripheral high-pass filters.

In summary, increasing filter level and size often modulated either mean fixation durations or saccade amplitudes. Thus, moderate attenuation of spatial frequencies affects temporal and spatial aspects of eye-movement behavior, but with a further increase of processing difficulty resources are often preserved by adapting either saccade timing or saccadic selection, but not both.

\section{Conclusions}
The present study demonstrates that eye-movement behavior during real-world scene viewing is impaired with increasing processing difficulty due to spatial frequency filters of varying type, size, and level in the central or the peripheral visual field. The adaptation to increasing processing difficulty is particularly evident in the modulation of saccade target selection (i.e., saccade amplitudes). Saccade timing is not necessarily adapted to processing difficulty---when visual-cognitive processing becomes too time-consuming, fixation durations are not prolonged any further, but instead, stimulus-independent timing is adapted. This finding is in good agreement with the assumption of a psychologically plausible trade-off relation between visual processing time and saccade generation.
 
\section{Acknowledgments}
This work was funded by Deutsche Forschungsgemeinschaft (grants LA 2884/1 to J.L. and EN 471/10 to R.E.). We thank Heiko Sch\"utt for helpful comments and Petra Schienmann and our student assistants for their help during data collection.


\bibliographystyle{elsarticle-harv}
\bibliography{references}

\appendix

\newpage
\section{Variance components and fixed effects estimated with linear mixed-effects models}

\begin{table}[htbp] 	
\caption{Variance components and fixed effects for fixation durations in Experiment 1. \emph{t}-values exceeding an absolute value of 1.96 were regarded as being statistically significant on the two-tailed 5\% level. \vspace{0.3cm}}
	\centering 
		\begin{tabular} {l r r r}
			\toprule       
			Variance components & SD $(\cdot 10^{-2})$ \\	                             
			\midrule
			Participants & 9.48 \\
			Scenes & 2.44  \\ 
			Residuals & 41.44 \\ \\
			\midrule
			Fixed effects & Coefficient & SE & t \\
						 & $(\cdot 10^{-2})$ & $(\cdot 10^{-3})$ \\
			\midrule  
			Unfiltered control low-pass vs. high-pass & 2.00  & 5.41 & 3.7 \\
			Unfiltered controls vs. all filter conditions & 4.37 & 2.36 & 18.5 \\
			Filtered control low-pass vs. high-pass & 2.35  & 5.51 & 4.3 \\
			Filtered controls vs. gaze-contingent filters & 1.12  &  2.44 & 4.6 \\
			Filter type & 2.29 & 3.64 &  6.3 \\
			Filter location & 1.56 & 1.84 & 8.5 \\
			Weak vs. moderate and strong filters (W-MS) & 2.20 & 1.94 & 11.3 \\
			Moderate vs. strong filter (M-S) & 0.90 & 2.26 & 4.0 \\
			Filter type $\times$ Filter location & 2.79 & 1.84 & 15.2 \\
			Filter type $\times$ W-MS & 1.83 & 1.94 & 9.4 \\
			Filter type $\times$ M-S & 0.84 & 2.26 & 3.7 \\
			Filter location $\times$ W-MS & 0.87 & 1.94 & 4.5 \\
			Filter location $\times$ M-S & -0.03 & 2.26 & -0.1 \\
			Filter type $\times$ Filter location $\times$ W-MS & 1.34 & 1.94 & 6.9 \\
			Filter type $\times$ Filter location $\times$ M-S & 0.57 & 2.25 & 2.5 \\ 
			\bottomrule
		\end{tabular}
\label{tab:LMM_FD_Exp1}
\end{table}

\begin{table}[htbp] 	
\caption{Variance components and fixed effects for saccade amplitudes in Experiment 1. \emph{t}-values exceeding an absolute value of 1.96 were regarded as being statistically significant on the two-tailed 5\% level. \vspace{0.3cm}}
	\centering 
		\begin{tabular} {l r r r}
			\toprule       
			Variance components & SD $(\cdot 10^{-2})$ \\	                             
			\midrule
			Participants & 5.57  \\
			Scenes & 2.39  \\ 
			Residuals & 25.45 \\ \\			
			\midrule
			Fixed effects & Coefficient & SE & t \\	
						& $(\cdot 10^{-2})$ & $(\cdot 10^{-3})$ \\
			\midrule  
			Unfiltered control low-pass vs. high-pass & -1.19  & 4.07 & -2.92 \\
			Unfiltered controls vs. all filter conditions & 0.16 & 1.42 & 1.11 \\
			Filtered control low-pass vs. high-pass & 2.36  & 4.12 & 5.73 \\
			Filtered controls vs. gaze-contingent filters & 1.34  &  1.47 & 9.08 \\
			Filter type & 1.31 & 3.28 &  3.98 \\
			Filter location & 8.14 & 1.11 & 73.54 \\
			Weak vs. moderate and strong filters (W-MS) & 0.007 & 1.17 & 0.06 \\
			Moderate vs. strong filter (M-S) & 0.27 & 1.36 & 1.96 \\
			Filter type $\times$ Filter location & 1.01 & 1.11 & 9.13 \\
			Filter type $\times$ W-MS & -0.65 & 1.17 & -5.53 \\
			Filter type $\times$ M-S & -0.06 & 1.36 & -0.47 \\
			Filter location $\times$ W-MS & -3.67 & 1.17 & -31.33 \\
			Filter location $\times$ M-S & -2.65 & 1.36 & -19.48 \\
			Filter type $\times$ Filter location $\times$ W-MS & -1.01 & 1.17 & -8.62 \\
			Filter type $\times$ Filter location $\times$ M-S & -0.74 & 1.36 & -5.45 \\
			\bottomrule
		\end{tabular}
\label{tab:LMM_SA_Exp1}
\end{table}

\begin{table}[htbp] 	
\caption{Variance components and fixed effects for fixation durations in Experiment 2. \emph{t}-values exceeding an absolute value of 1.96 were regarded as being statistically significant on the two-tailed 5\% level. \vspace{0.3cm}}		
	\centering 
		\begin{tabular} {l r r r}
			\toprule       
			Variance components & SD $(\cdot 10^{-2})$ \\
			\midrule
			Participants & 11.72  \\
			Scenes & 2.30  \\
			Residuals & 42.21 \\ \\			
			\midrule
			Fixed effects & Coefficient & SE & t \\	
						& $(\cdot 10^{-2})$ & $(\cdot 10^{-3})$ \\
			\midrule  
			Unfiltered control low-pass vs. high-pass & 1.30  & 5.51 & 2.36 \\
			Unfiltered controls vs. all filter conditions & 5.41 & 2.49 & 21.76 \\
			Filtered control low-pass vs. high-pass & 1.91  & 5.78 & 3.31 \\
			Filtered controls vs. gaze-contingent filters & 3.21 & 2.66 & 12.09 \\
			Filter type & 2.99 & 3.54 & 8.44 \\
			Filter location & 1.68 & 1.92 & 8.74 \\
			Small vs. medium and large filters (S-ML) & 3.77 & 2.02 & 18.64 \\
			Medium vs. large filter (M-L) & 4.69 & 2.37 & 19.77 \\
			Filter type $\times$ Filter location & 5.37 & 1.92 & 27.96 \\
			Filter type $\times$ S-ML & 0.36 & 2.02 & 1.80 \\
			Filter type $\times$ M-L & 0.78 & 2.37 & 3.30 \\
			Filter location $\times$ S-ML & 4.66 & 2.02 & 23.05 \\
			Filter location $\times$ M-L & 4.42 & 2.37 & 18.66 \\
			Filter type $\times$ Filter location $\times$ S-ML & 0.76 & 2.02 & 3.78 \\
			Filter type $\times$ Filter location $\times$ M-L & 0.25 & 2.37 & 1.04 \\
			\bottomrule
		\end{tabular}
\label{tab:LMM_FD_Exp2}
\end{table}

\begin{table}[htbp] 	
\caption{Variance components and fixed effects for saccade amplitudes in Experiment 2. \emph{t}-values exceeding an absolute value of 1.96 were regarded as being statistically significant on the two-tailed 5\% level. \vspace{0.3cm}}		
	\centering 
		\begin{tabular} {l r r r}
			\toprule       
			Variance components & SD $(\cdot 10^{-2})$ \\	                             
			\midrule
			Participants & 2.66  \\
			Scenes & 1.69  \\ 
			Residuals & 19.12 \\ \\			
			\midrule
			Fixed effects & Coefficient & SE & t \\	
						& $(\cdot 10^{-2})$ & $(\cdot 10^{-3})$ \\
			\midrule  
			Unfiltered control low-pass vs. high-pass & 0.89 & 3.00 & 2.96 \\
			Unfiltered controls vs. all filter conditions & 0.39 & 1.10 & 3.53 \\
			Filtered control low-pass vs. high-pass & 4.51 & 3.10 & 14.56 \\
			Filtered controls vs. gaze-contingent filters & 1.37 & 1.18 & 11.60 \\
			Filter type & 1.17 & 2.34 & 5.00 \\
			Filter location & 9.89 & 0.85 & 115.88 \\
			Small vs. medium and large filters (S-ML) & 1.02 & 0.90 & 11.37 \\
			Medium vs. large filter (M-L) & 0.44 & 1.05 & 4.15 \\
			Filter type $\times$ Filter location & 2.01 & 0.85 & 23.49 \\
			Filter type $\times$ S-ML & 1.02 & 0.90 & 11.32 \\
			Filter type $\times$ M-L & 0.66 & 1.05 & 6.24 \\
			Filter location $\times$ S-ML & 3.15 & 0.90 & 35.09 \\
			Filter location $\times$ M-L & 0.95 & 1.05 & 9.03 \\
			Filter type $\times$ Filter location $\times$ S-ML & 0.86 & 0.90 & 9.59 \\
			Filter type $\times$ Filter location $\times$ M-L & -0.002 & 1.05 & -0.02 \\	
			\bottomrule
		\end{tabular}		
\label{tab:LMM_SA_Exp2}
\end{table}

\end{document}